\documentclass[a4paper, 11pt]{article}\usepackage[]{graphicx}\usepackage[dvipsnames]{xcolor}
\makeatletter
\def\maxwidth{ %
  \ifdim\Gin@nat@width>\linewidth
    \linewidth
  \else
    \Gin@nat@width
  \fi
}
\makeatother

\definecolor{fgcolor}{rgb}{0.345, 0.345, 0.345}

\usepackage{framed}
\makeatletter
 {\par\unskip\endMakeFramed%
 \at@end@of@kframe}
\makeatother

\definecolor{shadecolor}{rgb}{.97, .97, .97}
\definecolor{messagecolor}{rgb}{0, 0, 0}
\definecolor{warningcolor}{rgb}{1, 0, 1}
\definecolor{errorcolor}{rgb}{1, 0, 0}
\newenvironment{knitrout}{}{} 

\usepackage{alltt}
\usepackage[T1]{fontenc}
\usepackage[utf8]{inputenc}
\usepackage[english]{babel}
\usepackage{graphics}
\usepackage[dvipsnames]{xcolor}
\usepackage{amsmath, amssymb}
\usepackage{doi} 
\usepackage[round]{natbib} 
\usepackage{booktabs} 
\usepackage{multirow} 
\usepackage[title]{appendix} 
\usepackage{nameref} 
\usepackage[onehalfspacing]{setspace} 
\usepackage[labelfont=bf,font=small]{caption} 

\usepackage{geometry}
\newcommand\longtitle{Power priors for replication studies}
\newcommand\shorttitle{Power priors for replication studies} 
\newcommand\subtitle{}
\newcommand\longauthors{Samuel Pawel\textsuperscript{$\star$}, Frederik Aust\textsuperscript{$\dagger$}, Leonhard Held\textsuperscript{$\star$}, Eric-Jan Wagenmakers\textsuperscript{$\dagger$}}
\newcommand\shortauthors{S. Pawel, F. Aust, L. Held, E.-J. Wagenmakers} 
\newcommand\affiliation{
  $\star$ Epidemiology, Biostatistics and Prevention Institute (EBPI), \\ Center for Reproducible Science (CRS), University of Zurich \\
  $\dagger$ Department of Psychological Methods, University of Amsterdam
}
\newcommand\mail{samuel.pawel@uzh.ch}
\title{
  \vspace{-3em}
  \textbf{\longtitle} \\
  \subtitle
}
\author{
  \textbf{\longauthors} \\
  \affiliation \\
  E-mail: \href{mailto:\mail}{\mail}
}
\date{September 19, 2023} 

\usepackage{hyperref}
\hypersetup{
  unicode=true,
  bookmarksopen=true,
  breaklinks=true,
  pdftitle={\shorttitle},
  pdfauthor={\shortauthors},
  pdfsubject={},
  pdfkeywords={},
  colorlinks=true,
  linkcolor=blue,
  anchorcolor=black,
  citecolor=blue,
  urlcolor=black,
}

\usepackage{fancyhdr}
\newcommand{\eg}{e.\,g.,\,} 
\DeclareMathOperator{\Nor}{N} 
\DeclareMathOperator{\Be}{Be} 
\DeclareMathOperator{\Var}{Var} 
\newcommand{\B}{\operatorname{{B}}} 
\newcommand{\given}{\,\vert\,} 
\DeclareMathOperator{\BF}{BF} 
\newcommand{\that}{\hat{\theta}} 

\newcommand{\h}[1]{\mathcal{H}_{#1}} 

\IfFileExists{upquote.sty}{\usepackage{upquote}}{}
\begin{document}
\maketitle

\vspace{-3em}
\begin{center}
  ~
\end{center}

\begin{center}
  \begin{minipage}{13cm} {\small
      \rule{\textwidth}{0.5pt} \\
      {\centering \textbf{Abstract} \\
        The ongoing replication crisis in science has increased interest in the
        methodology of replication studies. We propose a novel Bayesian analysis
        approach using power priors: The likelihood of the original study's data
        is raised to the power of $\alpha$, and then used as the prior
        distribution in the analysis of the replication data. Posterior
        distribution and Bayes factor hypothesis tests related to the power
        parameter $\alpha$ quantify the degree of compatibility between the
        original and replication study. Inferences for other parameters, such as
        effect sizes, dynamically borrow information from the original study.
        The degree of borrowing depends on the conflict between the two studies.
        The practical value of the approach is illustrated on data from three
        replication studies, and the connection to hierarchical modeling
        approaches explored. We generalize the known connection between normal
        power priors and normal hierarchical models for fixed parameters and
        show that normal power prior inferences with a beta prior on the power
        parameter $\alpha$ align with normal hierarchical model inferences using
        a generalized beta prior on the relative heterogeneity variance $I^2$.
        The connection illustrates that power prior modeling is unnatural from
        the perspective of hierarchical modeling since it corresponds to
        specifying priors on a relative rather than an absolute heterogeneity
        scale.
      } \\
      \rule{\textwidth}{0.5pt} \emph{Keywords}: Bayes factor, Bayesian
      hypothesis testing, Bayesian parameter estimation, hierarchical models,
      historical data}
  \end{minipage}
\end{center}

\section{Introduction}

Power priors form a class of informative prior distributions that allow data
analysts to incorporate historical data into a Bayesian analysis
\citep{Ibrahim2015}. The most basic version of the power prior is obtained by
updating an initial prior distribution with the likelihood of the historical
data raised to the power of $\alpha$, where $\alpha$ is usually restricted to
the range from zero (i.e., complete discounting) to one (i.e., complete
pooling). As such, the power parameter $\alpha$ specifies the degree to which
historical data are discounted, thereby providing a quantitative compromise
between the extreme positions of completely ignoring and fully trusting the
historical data.

\begin{sloppypar}
One domain where historical data are per definition available is the analysis of
replication studies. One pertinent question in this domain is the extent to
which a replication study has successfully replicated the result of an original
study \citep{NSF2019}. Many methods have been proposed to address this question
\citep[among others]{Bayarri2002, Verhagen2014, Johnson2016, Etz2016,
  vanAert2017, Ly2018, Hedges2019, Mathur2020, Held2020, Pawel2020, Pawel2020b,
  Held2021}. Here we propose a new and conceptually straightforward approach,
namely to construct a power prior for the data from the original study, and to
use that prior to draw inferences from the data of the replication study. The
power prior approach can accommodate two common notions of replication success:
First, the notion that the replication study should provide evidence for a
genuine effect. This can be quantified by estimating and testing an effect size
$\theta$, typically by assessing whether there is evidence that $\theta$ is
different from zero. Second, the notion that the data from the original and
replication studies should be compatible. This can be quantified by estimating
and testing of the power parameter $\alpha$. Values close to $\alpha = 1$
indicate compatibility as there is a complete pooling of both data sets, and
values close to $\alpha = 0$ indicate incompatibility as the original data are
completely discounted.
\end{sloppypar}

Below we first show how power priors can be constructed from data of an original
study under a meta-analytic framework (Section~\ref{sec:power-prior}). We then
shown how the power prior can be used for parameter estimation
(Section~\ref{sec:parameter-estimation}) and Bayes factor hypothesis testing
(Section~\ref{sec:hypothesis-testing}). Throughout, the methodology is
illustrated by application to data from three replication studies which were
part of a large-scale replication project \citep{Protzko2020}. In
Section~\ref{sec:hierarch}, we explore the connection to the alternative
hierarchical modeling approach for incorporating the original data
\citep{Bayarri2002, Bayarri2002b, Pawel2020}, which has been previously used for
evidence synthesis and compatibility assessment in replication settings. In
doing so, we identify explicit conditions under which posterior distributions
and tests can be reverse-engineered from one framework to the other.
Essentially, power prior inferences using the commonly assigned beta prior on
the power parameter $\alpha$ align with normal hierarchical model inferences if
either a generalized F prior is assigned to the between-study heterogeneity
variance $\tau^2$ which scales with the variance of the original data, or if a
generalized beta prior is assigned to the relative heterogeneity $I^2$. This
perspective also explains the observed difficulty of making conclusive
inferences about the power parameter $\alpha$, as it is difficult to make
inferences about a variance from two observations alone, and also because the
commonly assigned beta prior on $\alpha$ is entangled with the variance from the
data.

\section{Power prior modeling of replication studies}
\label{sec:power-prior}

Let $\theta$ denote an unknown effect size and $\that_{i}$ an estimate thereof
obtained from study $i \in \{o, r\}$ where the subscript indicates ``original''
and ``replication'', respectively. Assume that the likelihood of the effect
estimates can be approximated by a normal distribution
\begin{align*}
  \that_{i} \given \theta \sim \Nor(\theta, \sigma^{2}_{i})
\end{align*}
with $\sigma_{i}$ the (assumed to be known) standard error of the effect
estimate $\that_{i}$. The effect size may be adjusted for confounding variables,
and depending on the outcome variable, a transformation may be required for the
normal approximation to be accurate (\eg{} a log-transformation for an odds
ratio effect size). This is the same framework that is typically used in
meta-analysis, and it is applicable to many types of data and effect sizes
\citep[chapter 2.4]{Spiegelhalter2004}. There are, of course, situations where
the approximation is inadequate and modified distributional assumptions are
required (\eg{} for data from studies with small sample sizes and/or extreme
effect sizes).

The goal is now to construct a power prior for $\theta$ based on the data from
the original study. Updating of an (improper) flat initial prior
$f(\theta) \propto 1$ by the likelihood of the original data raised to a (fixed)
power parameter $\alpha$ leads to the normalized power prior
\begin{align}
  \theta \given \that_{o}, \alpha
  \sim \Nor\left(\that_{o}, \sigma^{2}_{o}/\alpha \right)
  \label{eq:pp}
\end{align}
as first proposed by \citet{Duan2005}, see also \citet{Neuenschwander2009}.
There are different ways to specify $\alpha$. The simplest approach fixes
$\alpha$ to an \emph{a priori} reasonable value, possibly informed by background
knowledge about the similarity of the two studies. Another option is to use the
empirical Bayes estimate \citep{Gravestock2017}, that is, the value of $\alpha$
that maximizes the likelihood of the replication data marginalized over the
power prior.
Finally, it is also possible to specify a prior distribution for $\alpha$, the
most common choice being a beta distribution $\alpha \given x, y \sim \Be(x, y)$
for a normalized power prior conditional on $\alpha$ as in~\eqref{eq:pp}. This
approach leads to a joint prior for the effect size $\theta$ and power parameter
$\alpha$ with density
\begin{align}
    f(\theta, \alpha \given \that_o, x, y)
    = \Nor(\theta \given \that_o, \sigma^2_o/\alpha) \, \Be(\alpha \given x, y)
    \label{eq:ppjoint}
\end{align}
where $\Nor(\cdot \given m, v)$ is the normal density function with mean $m$ and
variance $v$, and $\Be( \cdot \given x, y)$ is the beta density with parameters
$x$ and $y$. The uniform distribution ($x = 1$, $y = 1$) is often recommended as
the default choice \citep{Ibrahim2015}. We note that $\alpha$ does not have to
be restricted to the unit interval but could also be treated as a relative
precision parameter \citep{Held2017}. We will, however, not consider such an
approach since power parameters $\alpha > 1$ lead to priors with more
information than what was actually supplied by the original study.

\subsection{Parameter estimation}
\label{sec:parameter-estimation}

Updating the prior~\eqref{eq:ppjoint} with the likelihood of the replication
data leads to the posterior distribution
\begin{align}
  f(\alpha, \theta \given \that_{r}, \that_{o}, x, y)
  =& \frac{\Nor(\that_{r}\given \theta, \sigma^{2}_{r}) \,
     \Nor(\theta\given \that_{o}, \sigma^{2}_{o}/\alpha) \, \Be(\alpha\given x, y)}{
     f(\that_{r} \given \that_{o}, x, y)}.
     \label{eq:posterior}
\end{align}
The normalizing constant
\begin{align}
  f(\that_{r} \given \that_{o}, x, y)
  &= \int_{0}^{1}  \Nor(\that_{r}\given \that_{o},  \sigma^{2}_{r} + \sigma^{2}_{o}/\alpha)
  \, \Be(\alpha\given x, y) \, \text{d}\alpha
  \label{eq:normConst}
\end{align}
is generally not available in closed form but requires numerical integration
with respect to $\alpha$. If inference concerns only one parameter, a marginal
posterior distribution for either $\alpha$ or $\theta$ can be obtained by
integrating out the corresponding nuisance parameter from~\eqref{eq:posterior}.
In the case of the power parameter $\alpha$, this leads to
\begin{align}
  f(\alpha \given \that_{r}, \that_{o}, x, y)
  =& \frac{\Nor(\that_{r}\given \that_{o}, \sigma^{2}_{r} + \sigma^{2}_{o}/\alpha) \,
     \Be(\alpha\given x, y)}{f(\that_{r} \given \that_{o}, x, y)}
     \label{eq:margalpha}
\end{align}
whereas for the effect size $\theta$, this gives
\begin{align*}
  f(\theta \given \that_{r}, \that_{o}, x, y)
     &= \frac{\Nor(\that_{r}\given \theta, \sigma^{2}_{r})
     \, \mbox{B}(x + 1/2, y)}{
     f(\that_{r} \given \that_{o}, x, y) \, \sqrt{2\pi\sigma_o^2}
     \, \mbox{B}(x, y)}
     \, M\bigg\{x + 1/2, x + y + 1/2, -\frac{(\that_o - \theta)^2}{2\sigma^2_o}\bigg\}
\end{align*}
with
$\mbox{B}(z, w) = \int_0^1 t^{z-1}(1 - t)^{w-1} \,\text{d}t = \{\Gamma(z)\Gamma(w)\}/\Gamma(z + w)$
the beta function and
$M(a, b, z) = \{\int_0^1 \exp(zt) t^{a-1}(1-t)^{b-a-1} \,\text{d}t\}/\mbox{B}(b - a, a)$
the confluent hypergeometric function \citep[chapters 6 and 13]{Abramowitz1964}.

\subsubsection{Example ``Labels''}

We now illustrate the methodology on data from the large-scale replication
project by \citet{Protzko2020}. The project featured an experiment called
``Labels'' for which the original study reported the following conclusion:
\textit{``When a researcher uses a label to describe people who hold a certain
  opinion, he or she is interpreted as disagreeing with those attributes when a
  negative label is used and agreeing with those attributes when a positive
  label is used''} \citep[p. 17]{Protzko2020}. This conclusion was based on a
standardized mean difference effect estimate $\that_{o} = 0.21$
and standard error $\sigma_{o} = 0.05$ obtained from
$1577$ participants. Subsequently, four replication studies were
conducted, three of them by a different laboratory than the original one, and
all employing large sample sizes. Since the same original study was replicated
by three independent laboratories, this is an instance of a ``multisite''
replication design \citep{Mathur2020}. While in principle it would be possible
to analyze all of these studies jointly, we will show separate analyses for each
pair of original and replication study as it reflects the typical situation of
only one replication study being conducted per original study.
Section~\ref{sec:discussion} discusses possible extensions of the power prior
approach for joint analyses in multisite designs.

\begin{figure}[!htb]

\begin{knitrout}
\definecolor{shadecolor}{rgb}{0.969, 0.969, 0.969}\color{fgcolor}
\includegraphics[width=\maxwidth]{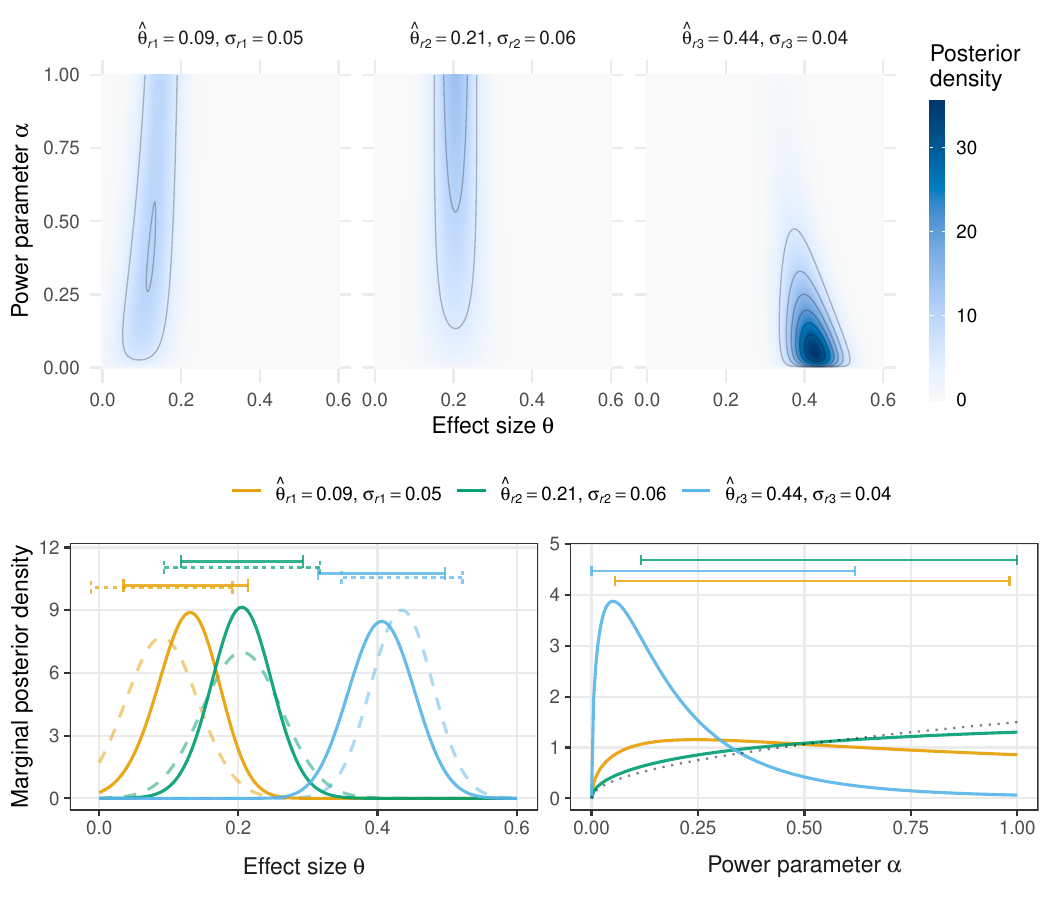} 
\end{knitrout}
\caption{Joint (top) and marginal (bottom) posterior distributions of effect
  size $\theta$ and power parameter $\alpha$ based on data from the ``Labels''
  experiment \citep{Protzko2020}. The dashed lines depict the posterior density
  for the effect size $\theta$ when the replication data are analyzed in
  isolation without incorporation of the original data. The horizontal error
  bars represent the corresponding 95\% highest posterior density credible
  intervals. The dotted line represents the limiting posterior density of the
  power parameter $\alpha$ for perfectly agreeing original and replication
  studies. }
\label{fig:post2d}
\end{figure}

Figure~\ref{fig:post2d} shows joint and marginal posterior distributions for
effect size $\theta$ and power parameter $\alpha$ based on the results of the
three external replication studies and a power prior for the effect size
$\theta$ constructed from the original effect estimate
$\that_{o} = 0.21$ (with standard error
$\sigma_{o} = 0.05$) and an initial flat prior
$f(\theta) \propto 1$. The power parameter $\alpha$ is assigned a uniform
$\Be(x = 1, y = 1)$ prior distribution. The
first replication found an effect estimate which was smaller than the original
one ($\that_{r1} = 0.09$ with
$\sigma_{r1} = 0.05$), whereas the other two replications
found effect estimates that were either identical
($\that_{r2} = 0.21$ with
$\sigma_{r2} = 0.04$) or larger
($\that_{r3} = 0.44$ with
$\sigma_{r3} = 0.06$) than that reported in the original
study. This is reflected in the marginal posterior distributions of the power
parameter $\alpha$, shown in the bottom right panel of Figure~\ref{fig:post2d}.
That is, the marginal distribution of the first replication (yellow) is slightly
peaked around $\alpha = 0.2$ suggesting some incompatibility with the original
study. In contrast, the second replication shows a marginal distribution (green)
which is monotonically increasing so that the value $\alpha = 1$ receives the
highest support, thereby indicating compatibility of the two studies. Finally,
the marginal distribution of the third replication (blue) is sharply peaked
around $\alpha = 0.05$ with 95\% credible interval from
$0$ to
$0.62$ indicating strong
conflict between this replication and the original study. The sharply peaked
posterior is in stark contrast to the relatively diffuse posteriors of the first
and second replications which hardly changed from the uniform prior. This is
consistent with the asymptotic behavior of normalized power priors identified in
\citet{Pawel2022c}; In case of data incompatibility, normalized power priors
with beta prior assigned to $\alpha$ permit arbitrarily peaked posteriors for
small values of $\alpha$. In contrast, for perfectly agreeing original and
replication studies ($\that_{o} = \that_{r}$) there is a limiting posterior for
$\alpha$ that gives only slightly more probability to values near one. The
limiting posterior is in this case a $\Be(3/2, 1)$ distribution, whose density
is indicated by the dotted line. One can see, that the (green) posterior from
the second replication is relatively close to the limiting posterior, despite
its finite sample size. Similarly, the corresponding (green) 95\% credible
interval from $0.12$ to
$1$ suggests that a
wide range of very low to very high $\alpha$ values remain credible despite the
excellent agreement of original and replication study.

The bottom left panel of Figure~\ref{fig:post2d} shows the marginal posterior
distribution of the effect size $\theta$. Shown is also the posterior
distribution of $\theta$ when the replication data are analyzed in isolation
(dashed line), to see the information gain from incorporating the original data
via a power prior. The degree of compatibility with the replication study
influences how much information is borrowed from the original study. For
instance, the (green) marginal posterior density based on the most compatible
replication ($\that_{r2} = 0.21$) is the most concentrated
among the three replications, despite the standard error being the largest
($\sigma_{r2} = 0.06$). Consequently, the 95\% credible
interval of $\theta$ is substantially narrower compared to the credible interval
from the analysis of the replication data in isolation (dashed green). In
contrast, the (blue) marginal posterior of the most conflicting estimate
($\that_{r3} = 0.44$) borrows less information and
consequently yields the least peaked posterior, despite the standard error being
the smallest ($\sigma_{r3} = 0.04$). In this case, the
conflict with the original study even inflates the variance of posterior
compared to the isolated replication posterior given by dashed blue line. This
is, for example, apparent through its 95\% credible interval
($0.31$ to
$0.5$) being even wider
than the credible interval
($0.35$ to
$0.52$) based on the
analysis of the replication data in isolation.

\subsection{Hypothesis testing}
\label{sec:hypothesis-testing}

In addition to estimating $\theta$ and $\alpha$, we may also be interested in
testing hypotheses about these parameters. Let $\h{0}$ and $\h{1}$ denote two
competing hypotheses, each of them with an associated prior
$f(\theta, \alpha \given \h{i})$ and a resulting marginal likelihood obtained
from integrating the likelihood of the replication data with respect to the
prior
\begin{align}
    \label{eq:marglikpow}
    f(\that_r \given \h{i}) = \int \Nor(\that_r \given \theta, \sigma^2_r) \,
    f(\theta, \alpha \given \h{i}) \,\text{d}\theta\,\text{d}\alpha
\end{align}
for $i \in \{0, 1\}$. A principled Bayesian hypothesis testing approach is to
compute the Bayes factor
\begin{align*}
  \BF_{01}(\that_r)
  = \frac{\Pr(\h{0} \given \that_r)}{\Pr(\h{1} \given \that_r)}
  \, \bigg/ \, \frac{\Pr(\h{0})}{\Pr(\h{1})}
  = \frac{f(\that_r \given \h{0})}{f(\that_r \given \h{1})}
\end{align*}
since it corresponds to the updating factor of the prior odds to the posterior
odds of the hypotheses based on the data $\that_{r}$ (first equality), or
because it represents the relative accuracy with which the hypotheses predict
the data $\that_{r}$ (second equality) \citep{Jeffreys1939, Good1958, Kass1995}.
A Bayes factor $\BF_{01}(\that_r) > 1$ provides evidence for $\h{0}$, whereas a
Bayes factor $\BF_{01}(\that_r) < 1$ provides evidence for $\h{1}$. The more the
Bayes factor deviates from one, the larger the evidence. In the following we
will examine the Bayes factors related to various hypotheses about $\theta$ and
$\alpha$.

\begin{sloppypar}
\subsubsection{Hypotheses about the effect size \texorpdfstring{$\boldsymbol{\theta}$}{}}

Researchers may be interested in testing the null hypothesis that there is no
effect ($\h{0} \colon \theta = 0$) against the alternative that there is an
effect ($\h{1} \colon \theta \neq 0$). We note that while the point null
hypothesis $\h{0}$ is often unrealistic, it is usually a good approximation to
more realistic interval null hypotheses that assign a distribution tightly
concentrated around zero \citep{Berger1987b, Ly2022}. Under $\h{0}$ there are no
free parameters, but under the alternative $\h{1}$ the specification of a prior
distribution for $\theta$ and $\alpha$ is required. A natural choice is to use
the normalized power prior based on the original data along with a beta prior
for the power parameter as in~\eqref{eq:ppjoint}. The associated Bayes factor is
then given by
\begin{align}
  \BF_{01}\{\that_{r}\given \h{1} \colon \alpha \sim \Be(x, y)\}
  &= \frac{f(\that_{r} \given \h{0} \colon \theta = 0)}{f\{\that_{r} \given \h{1}
    \colon \theta \given \alpha   \sim \Nor(\that_{o}, \sigma^{2}_{o}/\alpha),
    \alpha \sim \Be(x, y)\}} \nonumber \\
    &=  \frac{\Nor(\that_{r}\given 0, \sigma^{2}_{r})}{\int_{0}^{1} \Nor(\that_{r}\given \that_{o},
    \sigma^{2}_{r} + \sigma^{2}_{o}/\alpha) \, \Be(\alpha\given x, y) \, \text{d}\alpha}.
  \label{eq:bf01}
\end{align}
An intuitively reasonable choice for the prior of $\alpha$ under $\h{1}$ is a
uniform $\alpha \sim \Be(x=1, y=1)$ distribution. However, it is worth
noting that assigning a point mass $\alpha = 1$ leads to
\begin{align}
  \BF_{01}(\that_{r}\given \h{1} \colon \alpha = 1)
   = \frac{f(\that_{r} \given \h{0} \colon \theta = 0)}{f\{\that_{r} \given \h{1}
    \colon \theta \given \alpha  \sim \Nor(\that_{o}, \sigma^{2}_{o}/\alpha),
    \alpha = 1\}}
    =  \frac{\Nor(\that_{r}\given 0, \sigma^{2}_{r})}{ \Nor(\that_{r}\given \that_{o},
    \sigma^{2}_{o} + \sigma^{2}_{r})},
  \label{eq:bfr}
\end{align}
which is the \emph{replication Bayes factor} under normality
\citep{Verhagen2014, Ly2018, Pawel2020b}, that is, the Bayes factor contrasting
a point null hypothesis to the posterior distribution of the effect size based
on the original data (and in this case a uniform initial prior). A fixed
$\alpha = 1$ can also be seen as the limiting case of a beta prior with $y > 0$
and $x \to \infty$. The power prior version of the replication Bayes factor is
thus a generalization of the standard replication Bayes factor, one that allows
the original data to be discounted to some degree.
\end{sloppypar}

\subsubsection{Hypotheses about the power parameter \texorpdfstring{$\boldsymbol{\alpha}$}{}}

To quantify the compatibility between the original and replication study,
researchers may also be interested in testing hypotheses regarding the power
parameter $\alpha$. For example, we may want to test the hypothesis that the
data sets are ``compatible'' and should be completely pooled
($\h{\text{c}} \colon \alpha = 1$) against the hypothesis that they are
incompatible or ``different'' and the original data should be discounted to some
extent ($\h{\text{d}} \colon \alpha < 1$).

One approach is to assign a point prior $\h{\text{d}}\colon\alpha = 0$ which
represents the extreme position that the original data should be completely
discounted. This leads to the issue that for a flat initial prior
$f(\theta) \propto 1$, the power prior with $\alpha = 0$ is not proper and so
the resulting Bayes factor is only defined up to an arbitrary constant. Instead
of the flat prior, we may thus assign an uninformative but proper initial prior
to $\theta$, for instance, a unit-information prior
$\theta \sim \Nor(0, \kappa^{2})$ with $\kappa^{2}$ the variance from one
(effective) observation \citep{Kass1995b} as it encodes minimal prior
information about the direction or magnitude of the effect size
\citep{Best2021}. Updating the unit-information prior by the likelihood of the
original data raised to the power of $\alpha$ leads then to a
$\theta \given \alpha \sim \Nor\{\mu_{\alpha} = (\alpha\that_{o})/(\alpha + \sigma^2_{o}/\kappa^{2}), \sigma^{2}_{\alpha} = 1/(1/\kappa^{2} + \alpha/\sigma^{2}_{o})\}$
distribution, so the Bayes factor is
\begin{align}
  \BF_{\text{dc}}(\that_{r} \given \h{\text{d}}\colon\alpha = 0)
  = \frac{f\{\that_{r} \given \h{\text{d}}\colon \theta \given \alpha \sim
    \Nor(\mu_{\alpha}, \sigma^{2}_{\alpha}), \alpha = 0\}}{f\{\that_{r}
    \given \h{\text{c}}\colon \theta \given \alpha \sim \Nor(\mu_{\alpha}, \sigma^{2}_{\alpha}), \alpha = 1\}}
    = \frac{\Nor(\that_{r}\given 0, \sigma^{2}_{r} + \kappa^{2})}{\Nor(\that_{r}\given s
    \that_{o}, \sigma^{2}_{r} + s  \sigma^{2}_{o})}
    \label{eq:bfalpha}
\end{align}
with $s = 1 / (1 + \sigma^{2}_{o}/\kappa^{2})$.

An alternative approach that avoids the specification of a proper initial prior
for $\theta$ is to assign a prior to $\alpha$ under $\h{\text{d}}$. A suitable
class of priors is given by $\h{\text{d}} \colon \alpha \sim \Be(1, y)$ with
$y > 1$. The $\Be(1, y)$ prior has its highest density at $\alpha = 0$ and is
monotonically decreasing thus representing the more nuanced position that the
original data should only be partially discounted. The parameter $y$ determines
the extent of partial discounting and the simple hypothesis
$\h{\text{d}} \colon \alpha = 0$ can be seen as a limiting case when
$y \to \infty$. The resulting Bayes factor is given by
\begin{align}
  \BF_{\text{dc}}\{\that_{r}\given \h{\text{d}} \colon \alpha \sim \Be(1, y)\}
  &= \frac{f\{\that_{r} \given \h{\text{d}}\colon \theta \given \alpha \sim
    \Nor(\that_{o}, \sigma^{2}_{o}/\alpha), \alpha \sim \Be(1, y)\}}{f\{\that_{r}
  \given \h{\text{c}} \colon \theta \given \alpha \sim \Nor(\that_{o}, \sigma^{2}_{o}/\alpha), \alpha = 1\}}
  \nonumber \\
  &= \frac{\int_{0}^{1}\Nor(\that_{r}\given\that_{o}, \sigma^{2}_{r} + \sigma^{2}_{o}/\alpha) \,
    \Be(\alpha\given 1, y) \, \text{d}\alpha}{\Nor(\that_{r}\given\that_{o}, \sigma^{2}_{r}
  + \sigma^{2}_{o})}.
    \label{eq:bfdcrandom}
\end{align}

\subsubsection{Example ``Labels'' (continued)}

Table~\ref{tab:hypothesis} displays the results of the proposed hypothesis tests
applied to the three replications of the ``Labels'' experiment. The Bayes
factors contrasting $\h{0}\colon \theta = 0$ to $\h{1}\colon \theta \neq 0$ with
normalized power prior with uniform prior for the power parameter $\alpha$ under
the alternative (column
\mbox{$\BF_{01}\{\hat{\theta}_r \given \h{1} \colon \alpha \sim \Be(1, 1)\}$})
indicate neither evidence for absence nor presence of an effect in the first
replication, but decisive evidence for the presence of an effect in the second
and third replication. In all three cases, the Bayes factors are close to the
standard replication Bayes factors with $\alpha = 1$ under the alternative
(column $\BF_{01}(\hat{\theta}_r \given \h{1} \colon \alpha = 1)$).

\begin{table}[!htb]
  \centering
  \caption{Hypothesis tests for the replication studies of the ``Labels''
    experiment with original standardized mean difference effect estimate
    $\that_{o} = 0.21$ and standard error
    $\sigma_{o} = 0.05$. The columns indicate replication effect
    estimates $\that_{r}$, their standard errors $\sigma_{r}$, Bayes factors
    contrasting the absence of an effect $\h{0}\colon \theta = 0$ to the
    presence of an effect $\h{1} \colon \theta \neq 0$ with either a uniform
    prior $\alpha \sim \Be(x = 1, y = 1)$ or point prior $\alpha = 1$ under
    $\h{1}$, and Bayes factors contrasting study incompatibility
    $\h{\text{d}}\colon \alpha < 1$ to study compatibility
    $\h{\text{c}}\colon \alpha = 1$ with either complete discounting prior
    $\alpha = 0$ 
    or partial discounting prior $\alpha \sim \Be(1, y = 2)$ under
    $\h{\text{d}}$.}
  \label{tab:hypothesis}
\resizebox{1\textwidth}{!}{%
\begin{tabular}{ccccccc}
  \toprule & & & \multicolumn{2}{c}{Tests about the effect size $\theta$} & \multicolumn{2}{c}{Tests about the power parameter $\alpha$} \\ \cmidrule(lr){4-5} \cmidrule(lr){6-7} & $\hat{\theta}_r$ & $\sigma_r$ & $\BF_{01}\{\hat{\theta}_r \given \h{1} \colon \alpha \sim \Be(1, 1)\}$ & $\BF_{01}(\hat{\theta}_r \given \h{1} \colon \alpha = 1)$ & $\BF_{\text{dc}}(\hat{\theta}_r \given \h{\text{d}} \colon \alpha = 0)$ & $\BF_{\text{dc}}\{\hat{\theta}_r \given \h{\text{d}} \colon \alpha \sim \Be(1, 2)\}$ \\ 
  \midrule
  1 & 0.09 & 0.05 & 1/1.1 & 1.1 & 1/5.6 & 1.2 \\ 
    2 & 0.21 & 0.06 & 1/367 & 1/478 & 1/19 & 1/1.5 \\ 
    3 & 0.44 & 0.04 & < 1/1000 & < 1/1000 & 16 & 25 \\ 
   \bottomrule
\end{tabular}

}
\end{table}

In order to compute the Bayes factor for testing $\h{\text{d}}\colon \alpha = 0$
versus $\h{\text{c}}\colon \alpha = 1$ we need to specify a unit variance for
the unit-information prior. A crude approximation for the variance of a
standardized mean difference effect estimate is given by
$\Var(\that_i) = 4/n_{i}$ with $n_{i}$ the total sample size of the study, and
assuming equal sample size in both groups \citep[p. 5]{Hedges2021}. We may thus
set the variance of the unit-information prior to $\kappa^{2} = 2$ since a total
sample size of $n_{i} = 2$ (at least one observation from each group) is
required to estimate a standardized mean difference. Based on this choice, the
Bayes factors
$\BF_{\text{dc}}(\hat{\theta}_r \given \h{\text{d}}\colon \alpha = 0)$ in
Table~\ref{tab:hypothesis} indicate that the data provide substantial and strong
evidence for the compatibility hypothesis $\h{\text{c}}$ in the first and second
replication study, respectively, whereas the data indicate strong evidence for
complete incompatibility $\h{\text{d}}$ in the third replication study. The
Bayes factor
$\BF_{\text{dc}}\{\hat{\theta}_r \given \h{\text{d}}\colon \alpha \sim \Be(1, y = 2)\}$
in the right-most column with the partial discounting prior assigned under
hypothesis $\h{\text{d}}$ indicates absence of evidence for either hypothesis in
the first and second replication, but strong evidence for incompatibility
$\h{\text{d}}$ in the third replication. The apparent differences to the Bayes
factor with the complete discounting prior (column
$\BF_{\text{dc}}(\hat{\theta}_r \given \h{\text{d}}\colon \alpha = 0)$)
illustrate that in case of no conflict (study 2) or not too much conflict (study
1) the test with the partial discounting prior is less sensitive in diagnosing
(in)compatibility, but in case of substantial conflict (study 3) it is more
sensitive.

\begin{figure}[!htb]
\begin{knitrout}
\definecolor{shadecolor}{rgb}{0.969, 0.969, 0.969}\color{fgcolor}
\includegraphics[width=\maxwidth]{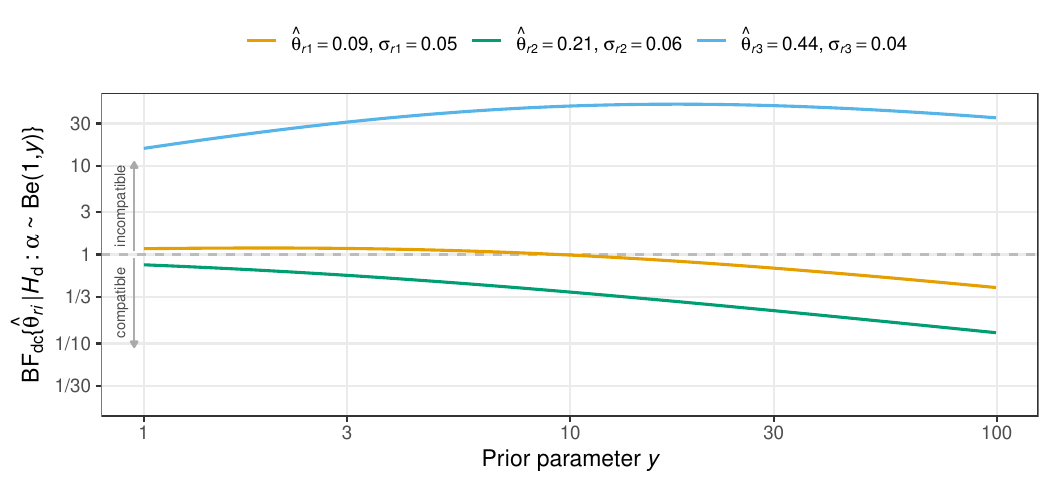} 
\end{knitrout}
\caption{Sensitivity of the Bayes factor $\BF_{\text{dc}}\{\hat{\theta}_r \given
  \h{\text{d}}\colon \alpha \sim \Be(1,
  y)\}$ with respect to the parameter $y$ of the partial discounting prior
  under $\h{\text{d}}$.}
\label{fig:sensitivityPrior}
\end{figure}

The previous analysis is based on a beta prior with $y = 2$ corresponding to a
linearly decreasing density in $\alpha$, Figure~\ref{fig:sensitivityPrior} shows
the Bayes factor for other values of $y$. We see that in the realistic range of
$y = 1$ (uniform prior) to $y = 100$ (almost all mass at $\alpha = 0$) the
results for the first and third replication hardly change, while for the second
replication the Bayes factor shifts from anecdotal evidence to stronger evidence
for compatibility.

To conclude, our analysis suggests that only the second replication was fully
successful in the sense that it provides evidence for the presence of an effect
while also being compatible with the original study. For the other two
replications the conclusions are more nuanced: In the first replication, there
is neither evidence for the absence nor the presence of an effect, but
substantial evidence for compatibility when a complete discounting prior is
used, and no evidence for (in)compatibility when a partial discounting prior is
used. Finally, in the third replication there is decisive evidence for an effect,
but also strong evidence of incompatibility with the original study.

\subsubsection{Bayes factor asymptotics}

Some of the Bayes factors in the previous example provided only modest evidence
for the test-relevant hypotheses despite the large sample sizes in original and
replication study. It is therefore of interest to understand the asymptotic
behavior of the proposed Bayes factors. For instance, we may wish to understand
what happens when the standard error of the replication study $\sigma_{r}$
becomes arbitrarily small (through an increase in sample size). Assume that
$\that_{r}$ is a consistent estimator of its true underlying effect size
$\theta_r$, so that as the standard error $\sigma_r$ goes to zero, the estimate
will converge in probability to the true effect size $\theta_r$. The true
replication effect size $\theta_{r}$ may be different from the true original
effect size $\theta_{o}$, for example, because the participant populations from
both studies systematically differ.

The limiting Bayes factors for testing the effect size $\theta$
from~\eqref{eq:bf01} and~\eqref{eq:bfr} are then given by
\begin{align*}
  \lim_{\sigma_{r} \downarrow 0}
  \BF_{\text{01}} \{\that_r \given \h{1} \colon \alpha \sim \Be(x, y)\}
    &= \frac{\delta(\theta_r) \, \sqrt{2\pi} \, \mbox{B}(x, y)}{\mbox{B}(x + 1/2, y)}
    \, M\bigg\{x+1/2, x + y + 1/2, -\frac{(\theta_r - \that_o)^2}{2\sigma^2_o}\bigg\}^{-1}
\end{align*}
and
\begin{align*}
    \lim_{\sigma_{r} \downarrow 0} \BF_{\text{01}} (\that_r \given \h{1} \colon \alpha = 1)
    &= \frac{\delta(\theta_r)}{\Nor(\theta_r \given \that_o, \sigma^2_o)},
\end{align*}
with $\delta(\cdot)$ the Dirac delta function. Both Bayes factors are hence
consistent \citep{Bayarri2012} in the sense that they indicate overwhelming
evidence for the correct hypothesis (i.e., the Bayes factors go to infinity/zero
if the true effect size $\theta_r$ is zero/non-zero). In contrast, the Bayes
factors for testing the power parameter $\alpha$ from~\eqref{eq:bfalpha}
and~\eqref{eq:bfdcrandom} converge to positive constants
\begin{align}
  \label{eq:boundsimple}
  \lim_{\sigma_{r} \downarrow 0} \BF_{\text{dc}} (\theta_r\given \h{\text{d}} \colon \alpha = 0) =
    \sqrt{1 - s} \, \exp\left[
    -\frac{1}{2} \, \left\{\frac{\theta_r^{2}}{\kappa^{2}} -
    \frac{(\theta_r - s\that_{o})^{2}}{s\sigma^{2}_{o}}\right\}\right]
\end{align}
and
\begin{align}
  \label{eq:boundcomposite}
  \lim_{\sigma_{r} \downarrow 0}
  \BF_{\text{dc}}\{\theta_r\given \h{\text{d}} \colon \alpha \sim \Be(1, y)\}
  &= \frac{\mbox{B}(3/2, y)}{\mbox{B}(1, y)} \,
  M\left\{y, y + 3/2, \frac{(\theta_r - \that_o)^2}{2\sigma^2_o}\right\}.
\end{align}
The amount of evidence one can find for either hypothesis thus depends on the
original effect estimate $\that_{o}$, the standard error $\sigma_{o}$, and the
true effect size $\theta_r$. For instance, in the ``Labels'' experiment we
have an original effect estimate $\that_{o} = 0.21$, a standard
error $\sigma_{o} = 0.05$, and a unit variance $\kappa^{2}=2$.
The bound~\eqref{eq:boundsimple} is minimized for a true effect size equal to
the original effect estimate $\theta_r = \that_{o} = 0.21$, so
the most extreme level we can obtain is
$\lim_{\sigma_{r} \downarrow 0} \BF_{\text{dc}} (\theta_r\given \h{\text{d}} \colon \alpha = 0) = 1/28$.
Similarly, the bound~\eqref{eq:boundcomposite} is minimized for
$\theta_r = \that_{o} = 0.21$ since then the confluent hypergeometric function
term becomes one, leading to
$\lim_{\sigma_{r} \downarrow 0} \BF_{\text{dc}} \{\theta_r\given\h{\text{d}} \colon \alpha \sim \Be(1, y = 2)\} = \mbox{B}(3/2, y)/\mbox{B}(1,y) = 1/1.9$.
Even in a perfectly precise replication study we cannot find more evidence, and
hence the posterior probability of $\h{\text{c}}\colon \alpha = 1$ cannot
converge to one.

While the Bayes factors~\eqref{eq:bfalpha} and~\eqref{eq:bfdcrandom} are
inconsistent if the replication data become arbitrarily informative, the
situation is different when also the original data become arbitrarily
informative (reflected by also the standard error $\sigma_o$ going to zero and
the original effect estimate $\that_o$ converging to its true effect size
$\theta_o$). The Bayes factor with $\h{\text{d}}\colon \alpha = 0$
from~\eqref{eq:bfalpha} is then consistent as the limit~\eqref{eq:boundsimple}
goes correctly to infinity/zero if the true effect size of the replication study
$\theta_r$ is different/equivalent from the true effect size of the original
study $\theta_o$. In contrast, the Bayes factor with
$\h{\text{d}} \colon \alpha \sim \Be(1, y)$ from~\eqref{eq:bfdcrandom} is still
inconsistent since it only shows the correct asymptotic behavior when the true
effect sizes are unequal (i.e., the Bayes factor goes to infinity) but not when
the effect sizes are equivalent, in which case it is still bounded by
$\mbox{B}(3/2, y)/\mbox{B}(1,y)$.

\subsubsection{Bayes factor design of replication studies}
\label{sec:design}
Now assume that the replication study has not yet been conducted and we wish to
plan for a suitable sample size. The design of replication studies should be
aligned with the planned analysis \citep{Anderson2017} and if multiple analyses
are performed, a sample size may be calculated that guarantees a sufficiently
conclusive analysis in each case \citep{Pawel2022d}.
In the power prior framework, samples size calculations may be based on either
hypothesis testing or estimation of the effect size $\theta$ or the power
parameter $\alpha$. Estimation based approaches have been developed by
\citet{Shen2023}. Here, we focus on samples size calculations based on Bayes
factor hypothesis testing as the methodology is still lacking.

\begin{sloppypar}
  In the case of testing the effect size $\theta$, \citet{Pawel2020b} studied
  Bayesian design of replication studies based on the Bayes
  factor~\eqref{eq:bfr} with $\alpha = 1$ under $\h{1}$, i.e., the replication
  Bayes factor under normality. They obtained closed-form expressions for the
  probability of replication success under $\h{0}$ and $\h{1}$ based on which
  standard Bayesian design can be performed \citep{Weiss1997,
    Gelfand2002,DeSantis2004,Schoenbrodt2017}. For the Bayes
  factor~\eqref{eq:bf01} with $\alpha \sim \Be(x, y )$ under $\h{1}$,
  closed-form expressions are not available
  anymore 
  and simulation or numerical integration have to be used for sample size
  calculations.
\end{sloppypar}

For tests related to the power parameter $\alpha$, there are also closed-form
expressions for the probability of replication success based on the Bayes
factor~\eqref{eq:bfalpha} with $\alpha = 0$ under $\h{\text{d}}$. We will now
show how these can be derived and used for determining the replication sample
size. With some algebra, one can show that
$\BF_{\text{dc}}(\that_{r} \given \h{\text{d}}\colon \alpha = 0) \leq \gamma$ is
equivalent to
\begin{align}
  \left\{\that_{r} -
  \frac{\that_o \, (\sigma^{2}_{r} + \kappa^{2})}{\kappa^2}\right\}^{2}
  \leq X  \label{eq:RScond}
\end{align}
with
\begin{align*}
    X =&
    \frac{(\sigma^{2}_{r} + \kappa^{2})(\sigma^{2}_{r} +
    s  \sigma^{2}_{o})}{\kappa^{2} - s  \sigma^{2}_{o}}
    \left\{\log\gamma^{2}- \log\left(\frac{\sigma^{2}_{r} + s  \sigma^{2}_{o}}{
            \sigma^{2}_{r} + \kappa^{2}}\right) - \frac{s^{2}  \that^{2}_{o}}{s  \sigma^{2}_{o} - \kappa^{2}}\right\}
\end{align*}
and $s = 1 / (1 + \sigma^{2}_{o}/\kappa^{2})$. Denote by $m_{i}$ and
$v_{i}$ the mean and variance of $\that_{r}$ under hypothesis
$i \in \{\text{d}, \text{c}\}$. The left hand side of~\eqref{eq:RScond} then
follows a scaled non-central chi-squared distribution under both hypotheses.
Hence the probability of replication success is given by
\begin{align}
  \Pr(\BF_{\text{dc}} \leq \gamma \given \h{i})
  &= \Pr\left(\chi^{2}_{1,\lambda_{i}} \leq X/v_{i} \right)
    \label{eq:PRS}
\end{align}
with non-centrality parameter
\begin{align*}
  \lambda_{i}
  = \left\{m_{i} -
  \frac{\that_o \, (\sigma^{2}_{r} + \kappa^{2})}{\kappa^2}
  \right\}^{2} \big / v_{i}.
\end{align*}

To determine the replication sample size, we can now use~\eqref{eq:PRS} to
compute the probability of replication success at a desired level $\gamma$ over
a grid of replication standard errors $\sigma_{r}$, and under either hypothesis
$\h{\text{d}}$ and $\h{\text{c}}$. The appropriate standard error $\sigma_r$ is
then chosen so that the probability for finding correct evidence is sufficiently
high under the respective hypothesis, and sufficiently low under the wrong
hypothesis. Subsequently, the standard error $\sigma_{r}$ needs to be translated
into a sample size, \eg{} for standardized mean differences via the
aforementioned approximation \mbox{$n_{r} \approx 4/\sigma^{2}_{r}$}.

\begin{figure}[!htb]
\begin{knitrout}
\definecolor{shadecolor}{rgb}{0.969, 0.969, 0.969}\color{fgcolor}
\includegraphics[width=\maxwidth]{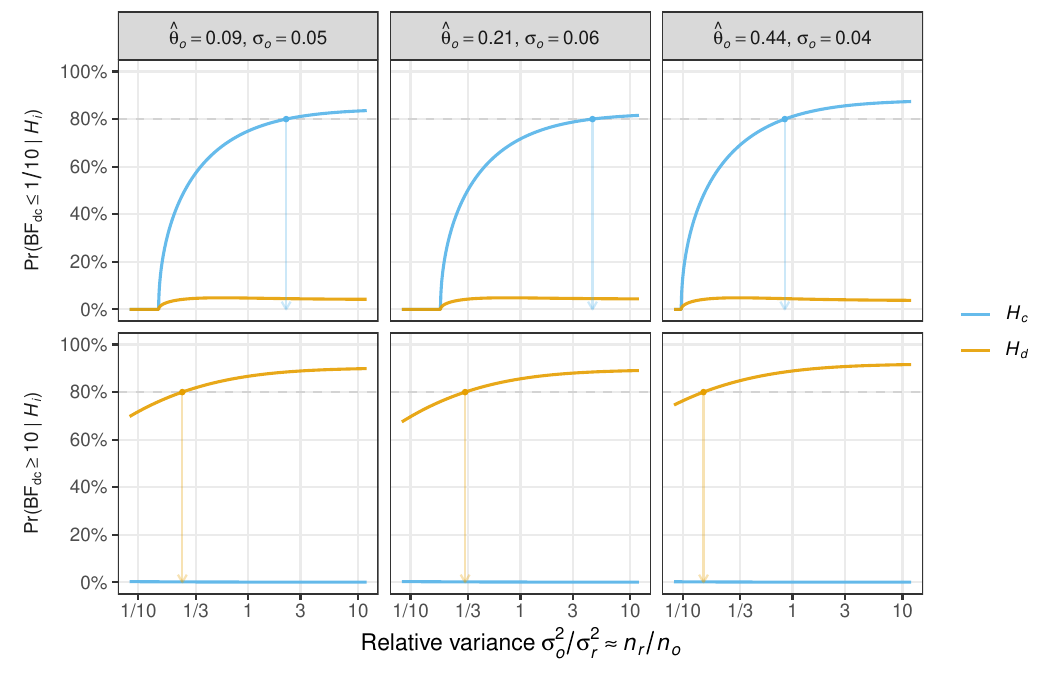} 
\end{knitrout}
\caption{Probability of replication success as a function of relative variance
  for the three replications of experiment ``Labels'' regarded as original
  study. The arrows point to the relative variance associated with an
  80\% probability under the respective hypotheses.}
\label{fig:ssd}
\end{figure}

\subsubsection{Example ``Labels'' (continued)}

Figure~\ref{fig:ssd} illustrates Bayesian design based on the Bayes factor
$\BF_{\text{dc}}(\that_{r} \given \h{\text{d}} \colon \alpha = 0)$ testing the
power parameter $\alpha$ from~\eqref{eq:bfalpha}. The three replication studies
from the experiment ``Labels'' are now regarded as original studies, and each
column of the figure shows the corresponding design of future replications. In
each plot, the probability for finding strong evidence for
$\h{\text{c}}\colon \alpha = 1$ (top) or $\h{\text{d}}\colon \alpha = 0$
(bottom) is shown as a function of the relative sample size. In both cases, the
probability is computed assuming that either $\h{\text{c}}$ (blue) or
$\h{\text{d}}$ (yellow) is true.

The curves look more or less similar for all three studies. We see from the
lower panels that the probability for finding strong evidence for $\h{\text{d}}$
is not much affected by the sample size of the replication study; it stays at
almost zero under $\h{\text{c}}$, while under $\h{\text{d}}$ it increases from
about 75\% to about 90\%. In contrast, the top panels show that the probability
for finding strong evidence for $\h{\text{c}}$ rapidly increases under
$\h{\text{c}}$ and seems to level off at an asymptote. Under $\h{\text{d}}$ the
probability stays below 5\% across the whole range.

The arrows in the plots display the required relative sample size to obtain
strong evidence with probability of $80\%$ under the
correct hypothesis. We see that original studies with smaller standard errors
require smaller relative sample sizes in the replication to achieve the same
probability of replication success. Under $\h{\text{c}}$ the required relative
sample sizes are larger than under $\h{\text{d}}$. However, while the
probability of misleading evidence under $\h{\text{c}}$ seems to be well
controlled under the determined sample size, under $\h{\text{d}}$ it stays
roughly 5\% for all three studies, and even for very large replication sample
sizes. Choosing the sample size based on finding strong evidence for
$\h{\text{c}}$ assuming $\h{\text{c}}$ is true thus also guarantees appropriate
error probabilities for finding strong evidence for $\h{\text{d}}$ in all three
studies. At the same time, it seems that the probability for finding misleading
evidence for $\h{\text{c}}$ cannot be reduced below around 5\% which might be
undesirably high for certain applications.

\section{Connection to hierarchical modeling of replication studies}
\label{sec:hierarch}

Hierarchical modeling is another approach that allows for the incorporation of
historical data in Bayesian analyses; moreover, hierarchical models have
previously been used in the replication setting \citep{Bayarri2002,
  Bayarri2002b, Pawel2020}. We will now investigate how the hierarchical
modeling approach is related to the power prior approach in the analysis of
replication studies, both in parameter estimation and hypothesis testing.

\subsection{Connection to parameter estimation in hierarchical models}

Assume a hierarchical model
\begin{subequations}
\label{eq:hierarch-model}
\begin{align}
  \that_{i} \given \theta_{i}\, &\sim \Nor(\theta_{i}, \sigma^{2}_{i}) \\
  \theta_{i} \given \theta_{*} &\sim \Nor(\theta_{*}, \tau^{2}) \\
  f(\theta_{*}) &\propto k
\end{align}
\end{subequations}
where for study $i \in \{o ,r\}$ the effect estimate $\that_i$ is normally
distributed around a study specific effect size $\theta_i$ which itself is
normally distributed around an overall effect size $\theta_{*}$. The
heterogeneity variance $\tau^2$ determines the similarity of the study specific
effect sizes $\theta_i$. The overall effect size $\theta_*$ is assigned an
(improper) flat prior $f(\theta_{*}) \propto k$, for some $k > 0$, which is a
common approach in hierarchical modeling of effect estimates \citep{Rover2021}.

\begin{figure}[!htb]
\begin{knitrout}
\definecolor{shadecolor}{rgb}{0.969, 0.969, 0.969}\color{fgcolor}
\includegraphics[width=\maxwidth]{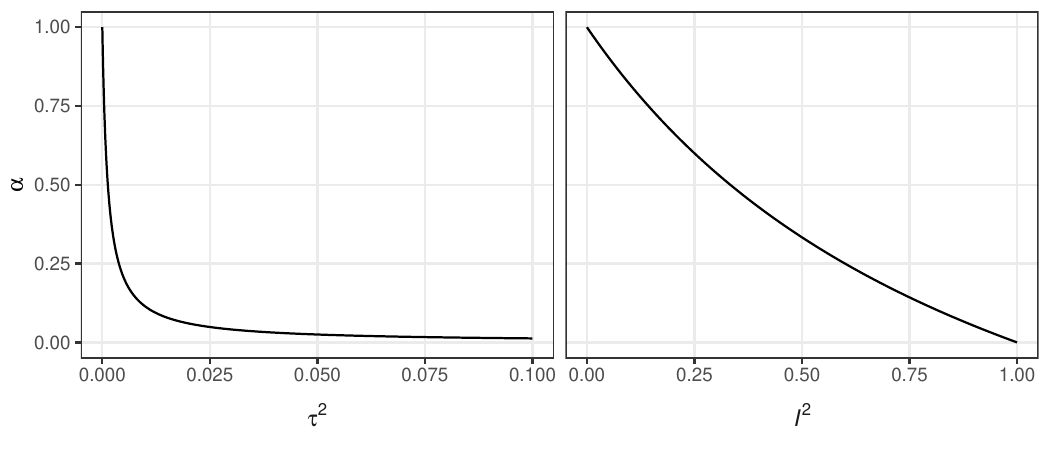} 
\end{knitrout}
\caption{The heterogeneity $\tau^{2}$ and relative heterogeneity
  $I^{2} = \tau^{2}/(\tau^{2} + \sigma^{2}_o)$ of a hierarchical model versus
  the power parameter $\alpha$ from a power prior model which lead to matching
  posteriors for the effect sizes $\theta$ and $\theta_{r}$. The variance of the
  original effect estimate $\sigma_{o}^2 = 0.05^{2}$ from the
  ``Labels'' experiment is used for the transformation to the heterogeneity
  scale $\tau^{2}$.}
\label{fig:I2}
\end{figure}

We show in Appendix~\ref{app:postHierarch} that under the hierarchical
model~\eqref{eq:hierarch-model} the marginal posterior distribution of the
replication specific effect size $\theta_{r}$ is given by
\begin{align}
  \label{eq:posthierarch}
  \theta_{r} \given \that_{o}, \that_{r}, \tau^{2}
  \sim \Nor\left(\frac{\that_{r}/\sigma^{2}_{r} + \that_{o}/(2\tau^{2} +
  \sigma^{2}_{o})}{1/\sigma^{2}_{r} + 1/(2\tau^{2} + \sigma^{2}_{o})},
  \frac{1}{1/\sigma^{2}_{r} + 1/(2\tau^{2} + \sigma^{2}_{o})}\right),
\end{align}
that is, a normal distribution whose mean is a weighted average of the
replication effect estimate $\that_r$ and the original effect estimate
$\that_o$. The amount of shrinkage of the replication towards the original
effect estimate depends on how large the replication standard error $\sigma_r$
is relative to the heterogeneity variance $\tau^2$ and the original standard
error $\sigma_o$. There exists a correspondence between the posterior for the
replication effect size $\theta_r$ from the hierarchical
model~\eqref{eq:posthierarch} and the posterior for the effect size $\theta$
under the power prior approach. Specifically, note that under the power prior
and for a fixed power parameter $\alpha$, the posterior of the effect size
$\theta$ is given by
\begin{align}
  \label{eq:postpower}
  \theta \given \that_{o}, \that_{r}, \alpha
  \sim \Nor\left(\frac{\that_{r}/\sigma^{2}_{r} +
  (\that_{o}\alpha)/\sigma^{2}_{o}}{1/\sigma^{2}_{r} + \alpha/\sigma^{2}_{o}},
  \frac{1}{1/\sigma^{2}_{r} + \alpha/\sigma^{2}_{o}}\right).
\end{align}
The hierarchical posterior ~\eqref{eq:posthierarch} and the power prior
posterior~\eqref{eq:postpower} thus match if and only if
\begin{align}
  \alpha = \frac{\sigma^{2}_{o}}{2\tau^{2} + \sigma^{2}_{o}},
  \label{eq:tau2alpha}
\end{align}
respectively
\begin{align}
  \tau^{2} = \left(\frac{1}{\alpha} - 1\right) \, \frac{\sigma^{2}_{o}}{2},
  \label{eq:alpha2tau}
\end{align}
which was first shown by \citet{Chen2006}. For instance, a power prior model
with $\alpha = 1$ corresponds to a hierarchical model with $\tau^{2} = 0$, and a
hierarchical model with $\tau^2 \to \infty$ corresponds to a power prior model
with $\alpha \downarrow 0$. In between these two extremes, however, $\alpha$ has
to be interpreted as a relative measure of heterogeneity since the
transformation to $\tau^2$ involves a scaling by the variance $\sigma^2_o$ of
the original effect estimate. For this reason, there is a direct correspondence
between $\alpha$ and the popular relative heterogeneity measure
$I^{2} = \tau^{2}/(\tau^{2} + \sigma^{2}_o)$ \citep{Higgins2002} computed from
$\tau^2$ and the variance of the original estimate $\sigma^2_o$, that is,
\begin{align*}
  \alpha = \frac{1 - I^{2}}{1 + I^{2}},
\end{align*}
with inverse of the same functional form. Figure~\ref{fig:I2} shows $\alpha$ and
the corresponding $\tau^2$ and $I^2$ values which lead to matching posteriors.

It has remained unclear whether or not a similar correspondence
exists in cases where $\alpha$ and $\tau^{2}$ are random and assigned prior
distributions. Here we confirm that there is indeed such a correspondence.
Specifically, the marginal posterior of the replication effect size $\theta_r$
from the hierarchical model matches with the marginal posterior of the effect
size $\theta$ from the power prior model if the prior density functions
$f_{\tau^2}(\cdot)$ and $f_{\alpha}(\cdot)$ of $\tau^2$ and $\alpha$ satisfy
\begin{align}
  \label{eq:matchCond}
  f_{\tau^2}(\tau^{2}) = f_{\alpha}\left(\frac{\sigma^{2}_{o}}{2 \tau^{2} + \sigma^{2}_{o}}\right) \,
  \frac{2\sigma^{2}_{o}}{(2 \tau^{2} + \sigma^{2}_{o})^{2}}
\end{align}
for every $\tau^2 \geq 0$, see Appendix~\ref{app:mapping} for details.
Importantly, the correspondence condition~\eqref{eq:matchCond} involves a
scaling by the variance from the original effect estimate $\sigma^2_o$, meaning
that also in this case $\alpha$ acts similar to a relative heterogeneity
parameter. This can also be seen from the correspondence condition between
$\alpha$ and $I^2 = \tau^2/(\sigma^2_o + \tau^2)$, which can be derived in
exactly the same way as the correspondence between $\alpha$ and $\tau^2$. That
is, the marginal posteriors of $\theta$ and $\theta_r$ match if the prior
density functions $f_{I^2}(\cdot)$ and $f_{\alpha}(\cdot)$ of $I^2$ and $\alpha$
satisfy
\begin{align}
  \label{eq:matchCondI2}
  f_{I^2}(I^{2}) = f_{\alpha}\left(\frac{1 - I^2}{1 + I^2}\right) \frac{2}{(1 + I^2)^2}
\end{align}
for every $0 \leq I^2 \leq 1$.

Interestingly, conditions~\eqref{eq:matchCondI2} and~\eqref{eq:matchCond} imply
that a beta prior on the power parameter $\alpha \sim \Be(x, y)$ corresponds to
a generalized F prior on the heterogeneity
$\tau^2 \sim \mbox{GF}(y, x, 2/\sigma^2_o)$ and a generalized beta prior on the
relative heterogeneity $I^2 \sim \mbox{GBe}(y, x, 2)$, see
Appendix~\ref{app:distribution} for details on both distributions. This
connection provides a convenient analytical link between hierarchical modeling
and the power prior framework, as beta priors for $\alpha$ are almost
universally used in applications of power priors. The result also illustrates
that the power prior framework seems unnatural from the perspective of
hierarchical modeling since it corresponds to specifying priors on the $I^2$
scale rather than on the $\tau^2$ scale. The same prior on $I^2$ will imply
different degrees of informativeness on the $\tau^2$ scale for original effect
estimates $\that_o$ with different variances $\sigma^2_o$ since $I^2$ is
entangled with the variance of the original effect estimate.

\begin{figure}[!htb]
\begin{knitrout}
\definecolor{shadecolor}{rgb}{0.969, 0.969, 0.969}\color{fgcolor}
\includegraphics[width=\maxwidth]{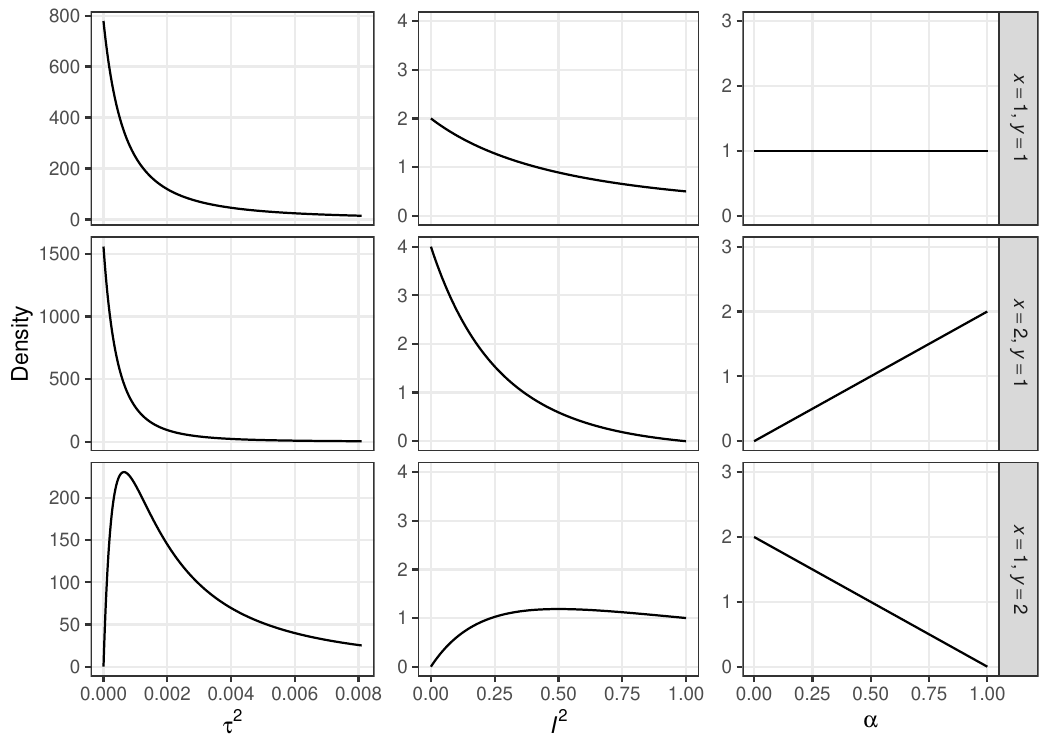} 
\end{knitrout}

\caption{Priors on the heterogeneity $\tau^2 \sim \mbox{GF}(y, x, 2/\sigma^2_o)$
(left), the relative heterogeneity
$I^2 = \tau^2/(\sigma^2_o + \tau^2) \sim \mbox{GBe}(y, x, 2)$ (middle) and the
power parameter $\alpha \sim \Be(x, y)$ (right) that lead to matching marginal
posteriors for effect sizes $\theta$ and $\theta_{r}$. The variance of the
original effect estimate $\sigma_{o}^2 = 0.05^{2}$ from
the ``Labels'' experiment is used for the transformation to the
heterogeneity scale $\tau^{2}$.}
\label{fig:matchingpriors}
\end{figure}

Figure~\ref{fig:matchingpriors} provides three examples of matching priors using
the variance of the original effect estimate from the ``Labels'' experiment for
the transformation to the heterogeneity scale $\tau^2$.
The top row of Figure~\ref{fig:matchingpriors} shows that the uniform prior on
$\alpha$ corresponds to a
$f(\tau^{2}) \propto \sigma^{2}_{o}/(2\tau^{2} + \sigma^{2}_{o})^{2}$ prior
which is similar to the ``uniform shrinkage'' prior
$f(\tau^{2}) \propto \sigma^{2}_{o}/(\tau^{2} + \sigma^{2}_{o})^{2}$
\citep{Daniels1999}. This prior has the highest density at $\tau^{2} = 0$ but
still gives some mass to larger values of $\tau^{2}$. Similarly, on the scale of
$I^2$ the prior slightly favors smaller values. The middle row of
Figure~\ref{fig:matchingpriors} shows that the $\alpha \sim \Be(2, 1)$ prior
---indicating more compatibility between original and replication than the
uniform prior--- gives even more mass to small values of $\tau^{2}$ and $I^2$,
and also has the highest density at $\tau^2 = 0$ and $I^2 = 0$. In contrast, the
bottom row of Figure~\ref{fig:matchingpriors} shows that the
$\alpha \sim \Be(1, 2)$ prior ---indicating less compatibility between original
and replication than the uniform prior--- gives less mass to small $\tau^{2}$
and $I^2$, and has zero density at $\tau^{2} = 0$ and $I^2 = 0$.

\subsection{Connection to hypothesis testing in hierarchical models}

Two types of hypothesis tests can be distinguished in the hierarchical model;
tests for the overall effect size $\theta_*$ and tests for the heterogeneity
variance $\tau^2$. In all cases, computations of marginal likelihoods of the
form
\begin{align}
    \label{eq:marglikhierarch}
    f(\that_r \given \h{i}) = \int \Nor(\that_r \given \theta_*, \sigma^2_r + \tau^2) \,
    f(\theta_*, \tau^2 \given \h{i})
    \, \text{d}\theta_* \, \text{d}\tau^2
\end{align}
with $i \in \{j, k\}$ are required for obtaining Bayes factors
$\BF_{jk}(\that_r) = f(\that_r \given \h{j})/f(\that_r \given \h{k})$ which
quantify the evidence that the replication data $\that_r$ provide for a
hypothesis $\h{k}$ over a competing hypothesis $\h{j}$. Under each hypothesis a
joint prior for $\tau^2$ and $\theta_*$ needs to be assigned.

As with parameter estimation, it is of interest to investigate whether there is
a correspondence with hypothesis tests from the power prior framework from
Section~\ref{sec:hypothesis-testing}. For two tests to match, one needs to
assign priors to $\tau^2$ and $\theta_*$, respectively, to $\alpha$ and $\theta$
so that the marginal likelihood~\eqref{eq:marglikhierarch} equals the marginal
likelihood from the power prior model~\eqref{eq:marglikpow} under both
test-relevant hypotheses.

Concerning the generalized replication Bayes factor from~\eqref{eq:bf01} testing
$\h{0} \colon \theta = 0$ versus $\h{1} \colon \theta \neq 0$, one can show that
it matches with the Bayes factor contrasting $\h{0} \colon \theta_* = 0$ versus
$\h{1} \colon \theta_* \neq 0$ with
\begin{align*}
  &\h{0}\colon \theta_{*} = 0& &\text{versus}&
  \h{1}\colon \theta_{*} \given \tau^{2} &\sim \Nor(\that_{o}, \sigma^{2}_{o} + \tau^{2}) \\
  &\phantom{\h{0}\colon} \tau^{2} = 0& &&
  \tau^2 &\sim \mbox{GF}(y, x, \sigma^2_o/2)
\end{align*}
for the replication data in in the hierarchical framework. The Bayes factor thus
compares the likelihood of the replication data under the hypothesis $\h{0}$
postulating that the global effect size $\theta_{*}$ is zero and that there is
no effect size heterogeneity, relative to the likelihood of the data under the
hypothesis $\h{1}$ postulating that $\theta_{*}$ follows the posterior based on
the original data and an initial flat prior for $\theta_{*}$ along with a
generalized F prior on the heterogeneity $\tau^2$. Setting the heterogeneity to
$\tau^2 = 0$ under $\h{1}$ instead produces the replication Bayes factor under
normality from~\eqref{eq:bfr}.

The Bayes factor~\eqref{eq:bfalpha} that tests complete discouting
$\h{\text{d}}\colon \alpha = 0$ versus complete compatibility
$\h{\text{c}}\colon \alpha = 1$ can be obtained in the hierarchical framework by
contrasting
\begin{align*}
  &\h{\text{d}}\colon \theta_{*} \sim \Nor(0, \kappa^{2})&
  &\text{versus}&
  &\h{\text{c}} \colon \theta_{*} \sim \Nor(s\,\that_{o}, s\,\sigma^{2}_{o})& \\
  &\phantom{\h{\text{d}}\colon} \tau^{2} = 0& &&
  &\phantom{\h{\text{c}}\colon} \tau^{2} = 0&
\end{align*}
with 
$s = 1 / (1 + \sigma^{2}_{o}/\kappa^{2})$.
Hence, the Bayes factor compares the likelihood of the replication data under
the initial unit-information prior relative to the likelihood of the replication
data under the unit-information prior updated by the original data, assuming no
heterogeneity under either hypothesis (so that the hierarchical model collapses
to a fixed effects model). Although this particular test relates to the power
parameter $\alpha$ in the power prior model, it is surprisingly unrelated to
testing the heterogeneity variance $\tau^2$ in the hierarchical model.

The Bayes factor~\eqref{eq:bfdcrandom} testing $\h{\text{d}}\colon \alpha < 1$
versus $\h{\text{c}}\colon \alpha = 1$ using the partial discounting prior
$\h{\text{d}} \colon \alpha \sim \Be(1, y)$ corresponds to testing
$\h{\text{d}}\colon \tau^2 > 0$ versus $\h{\text{c}}\colon \tau^2 = 0$ with
priors
\begin{align*}
  \h{\text{d}}\colon \theta_{*} \given \tau^{2} &\sim \Nor(\that_{o}, \sigma^{2}_{o} + \tau^{2})
  &\text{versus}& &
  \h{\text{c}} \colon \theta_{*} \given \tau^{2} &\sim \Nor(\that_{o}, \sigma^{2}_{o} + \tau^{2}) \\
  \tau^2  &\sim \mbox{GF}(y, 1, \sigma^2_o/2) &&&
  \tau^{2}&= 0
\end{align*}
The test for compatibility via the power parameter $\alpha$ is thus equivalent
to a test for compatibility via the heterogeneity $\tau^2$ (to which a
generalized F prior is assigned) after updating of a flat prior for $\theta_*$
with the data from the original study.

\subsection{Bayes factor asymptotics in the hierarchical model}

\begin{figure}[!htb]
\begin{knitrout}
\definecolor{shadecolor}{rgb}{0.969, 0.969, 0.969}\color{fgcolor}
\includegraphics[width=\maxwidth]{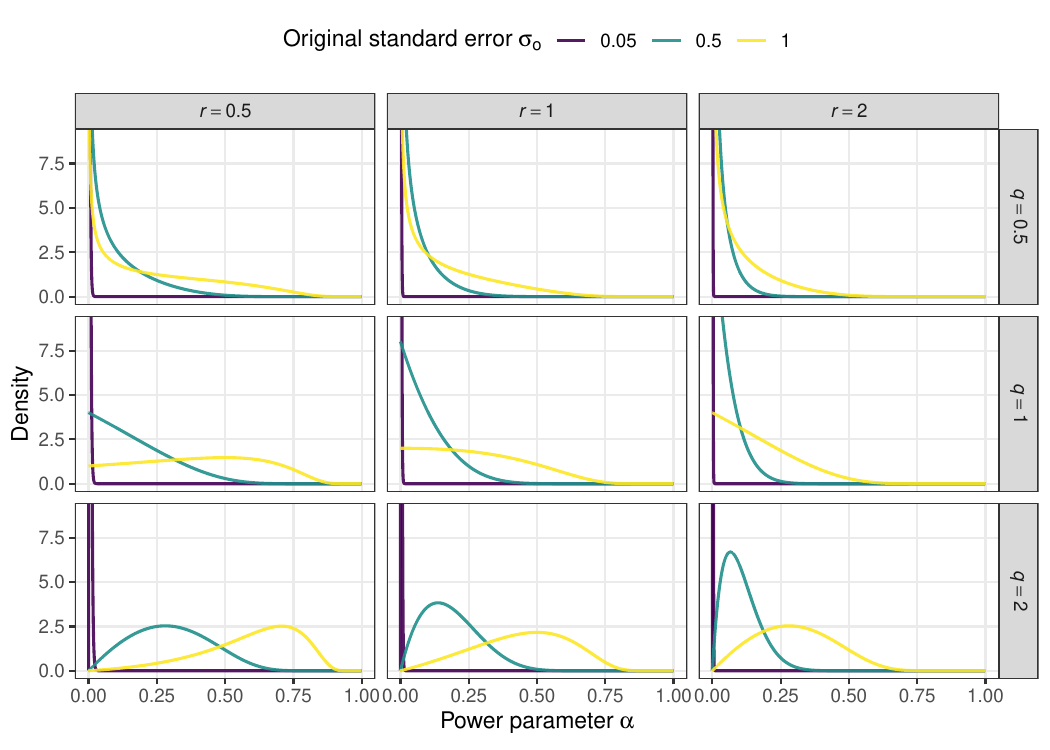} 
\end{knitrout}
\caption{Prior for the power parameter $\alpha$ implied by an inverse gamma
  prior $\h{\text{d}} \colon \tau^2 \sim \mbox{IG}(q, r)$ in a hierarchical
  model with consistent test for $\h{\text{c}}\colon \tau^{2} = 0$ versus
  $\h{\text{d}}\colon \tau^{2} > 0$.}
\label{fig:priorconsistent}
\end{figure}

Like the original test of $\h{\text{c}}\colon \alpha = 1$ versus
$\h{\text{d}}\colon \alpha \sim \Be(1, y)$, the corresponding test of $\tau^2$
is inconsistent in the sense that when the standard errors from both studies go
to zero ($\sigma_o \downarrow 0$ and $\sigma_r \downarrow 0$) and their true
effect sizes are equivalent ($\theta_o = \theta_r$), the Bayes factor
$\BF_{\text{dc}}$ does not go to zero (to indicate overwhelming evidence for
$\h{\text{c}}\colon \tau^2 = 0$) but converges to a positive constant. It is,
however, possible to construct a consistent test for
$\h{\text{c}}\colon \tau^2 = 0$ when we assign a different prior to $\tau^2$
under $\h{\text{d}}\colon \tau^2 > 0$. For instance, when we assign an inverse
gamma prior $\h{\text{d}} \colon \tau^2 \sim \mbox{IG}(q, r)$ with shape $q$ and
scale $r$, the Bayes factor is given by
\begin{align*}
    \BF_{\text{dc}}\{\that_r \given \h{\text{d}} \colon \tau^2 \sim \mbox{IG}(q, r)\}
    = \frac{\int \Nor(\that_r \given \that_o, \sigma^2_r + \sigma^2_o + 2\tau^2) \,
    \mbox{IG}(\tau^2 \given q, r) \, \text{d}\tau^2}{\Nor(\that_r \given \that_o,
    \sigma^2_r + \sigma^2_o)}
\end{align*}
with $\mbox{IG}(\cdot \given q, r)$ the density function of the inverse gamma
distribution. The limiting Bayes factor is therefore
\begin{align*}
    \lim_{\sigma_o, \sigma_r \downarrow 0} \BF_{\text{dc}}\{\that_r \given \h{\text{d}} \colon \tau^2 \sim \mbox{IG}(q, r)\}
    = \frac{\Gamma(q + 1/2)\{r + (\theta_r - \theta_o)^2/4\}^{-(q + 1/2)}}{\delta(\theta_r - \theta_o) \, \sqrt{4\pi}},
\end{align*}
so it correctly goes to zero/infinity when the effect sizes $\theta_r$ and
$\theta_o$ are equivalent/different. To understand why the test with
$\h{\text{d}} \colon \tau^2 \sim \mbox{IG}(q,r)$ is consistent, but the original
test with \mbox{$\h{\text{d}} \colon \alpha \sim \Be(1, y)$} is not, one can
transform the consistent test on $\tau^2$ to the corresponding test on $\alpha$.
The inverse gamma prior for $\tau^2$ implies a prior for $\alpha$ with density
\begin{align}    \label{eq:igalpha}
    f(\alpha \given q, r)
    &= \frac{r^q}{\Gamma(q)} \, \frac{\alpha^{q - 1}}{(1 - \alpha)^{q + 1}} \,
    \left(\frac{2}{\sigma^2_o}\right)^{q} \,
    \exp\left\{-\frac{2 \,r\, \alpha}{\sigma^2_o(1 - \alpha)}\right\}.
\end{align}
The Bayes factor contrasting $\h{\text{c}}\colon \alpha = 1$ versus
$\h{\text{d}}\colon \alpha < 1$ with prior~\eqref{eq:igalpha} assigned to
$\alpha$ under $\h{\text{d}}$ will thus produce a consistent test. The prior is
shown in Figure~\ref{fig:priorconsistent} for different parameters $q$ and $r$
and original standard errors $\sigma_{o}$. We see that the prior depends on the
standard error of the original effect estimate $\sigma_o$, the smaller
$\sigma_{o}$ the more the prior is shifted towards zero. For example, the
standard error $\sigma_{o} = 0.05$ from the ``Labels''
experiment leads to priors that are almost indistinguishable from a point mass
at $\alpha = 0$.
The prior thus ``unscales'' $\alpha$ from the
original standard error $\sigma_o$, thereby leading to a consistent test for
study compatibility and resolving the inconsistency property of the beta prior.

\section{Discussion}
\label{sec:discussion}

We showed how the power prior framework can be used for design and analysis of
replication studies. The approach supplies analysts with a suite of methods for
assessing effect sizes and study compatibility. Both aspects can be tackled from
an estimation or a hypothesis testing perspective, and the choice between the
two is primarily philosophical. We believe that both perspectives provide
valueable inferences that complement each other. Visualizations of joint and
marginal posterior distributions are highly informative in terms of the
available uncertainty. However, the power parameter $\alpha$ is an abstract
quantity disconnected from actual scientific phenomena. Testing hypotheses of
complete discounting versus complete pooling may therefore be more intuitive for
researchers. Both approaches also suffer from similar problems: If the original
and replication data are in perfect agreement, the posterior distribution of
$\alpha$ hardly changes from the prior. For example, for the commonly used
uniform prior $\alpha \sim \mathrm{Be}(x = 1, y = 1)$, we can at best obtain a
$\alpha \,\vert\, \hat{\theta}_{r} \sim \mathrm{Be}(x + 1/2 = 3/2, y = 1)$
posterior \citep{Pawel2022c}. This means that for a ``compatibility threshold''
of, say, 0.8, we can never have a posterior probability higher than
$\mathrm{Pr}(\alpha > 0.8 \,\vert\, \hat{\theta}_{r}) =
0.28$,
and for a threshold of 0.9 it is even lower
$\mathrm{Pr}(\alpha > 0.9 \,\vert\, \hat{\theta}_{r}) =
0.15$.
The fact that the Bayes factor for testing
$\h{\text{d}}\colon \alpha \sim \Be(1, y)$ against
$\h{\text{c}}\colon \alpha = 1$ is inconsistent, i.e., bounded from below by
a positive constant $\B(3/2,y)/\B(1,y)$, simply presents the problem from a
different perspective.

We also showed how the power prior approach is connected to hierarchical
modeling, and gave conditions under which posterior distributions and hypothesis
tests correspond between normal power prior models and normal hierarchical
models. This connection provides an intuition for why even with highly precise
and compatible original and replication study one can hardly draw conclusive
inferences about the power parameter $\alpha$; the power parameter $\alpha$ has
a direct correspondence to the relative heterogeneity variance $I^2$, and an
indirect correspondence to the heterogeneity variance $\tau^2$ in a hierarchical
model. Making inferences about a heterogeneity variance from two studies alone
seems like a virtually impossible task since the ``unit of information'' is the
number of studies and not the number of samples within a study. Moreover, Bayes
factor hypothesis tests related to $\alpha$ have the undesirable asymptotic
property of inconsistency if a beta prior is assigned to $\alpha$. This is
because the prior scales with the variance of the original data, just as a beta
prior for $I^2$ would in a hierarchical model. The identified link may also have
computational advantages, e.g., it may be possible to estimate power prior
models using the hierarchical model estimation procedures, or vice versa, but
more research is needed on the connection in more complex situations that depart
from normality assumptions.

Which of the two approaches should data analysts use in practice? We believe
that the choice should be primarily guided by whether the hierarchical or the
power prior model is \emph{scientifically} more suitable for the studies at
hand. If data analysts deem it scientifically plausible that the studies'
underlying effect sizes are connected via an overarching distribution then the
hierarchical model may be more suitable, particularly because the approach
naturally generalizes to more than two studies. On the other hand, if data
analysts simply want to downweight the original studies' contribution depending
on the observed conflict, the power prior approach might be more suitable. The
identified limitations for inferences related to the power parameter $\alpha$
should, however, be kept in mind when beta priors are assigned to the power
parameter $\alpha$.

There are also situations where the hierarchical and power prior frameworks can
be combined, for example, when multiple replications of a single original study
are conducted (multisite replications). In that case, one may model the
replication effect estimates in a hierarchical fashion but link their overall
effect size to the original study via a power prior. Multisite replications are
thus the opposite of the usual situation in clinical trials where several
historical ``original'' studies but only one current ``replication'' study is
available \citep{Gravestock2019}.

Another commonly used Bayesian approach for incorporating historical data are
\emph{robust mixture priors}, i.e., priors which are mixtures of the posterior
based on the historical data and an uninformative prior distribution
\citep{Schmidli2014}. We conjecture that inferences based on robust mixture
priors can be reverse-engineered within the framework of power priors through
Bayesian model averaging over two hypotheses about the power parameter; however,
more research is needed to explore the relationship between the two approaches.

The proposed methods are based on the standard meta-analytic assumption of
approximate normality of effect estimates with known variances. This makes our
methodology applicable to a wide range of effect sizes that may arise from
different data models. However, in some situations this assumption may be
inadequate, for example, when studies have small sample sizes. In this case, the
methods could be modified to use the exact likelihood of the data (\eg{}
binomial or $t$), as in \citet{Bayarri2002}, who used a $t$ likelihood. However,
the methodology would need to be adapted for each effect size type. Therefore,
future work may examine specific data models in more detail to obtain more
precise inferences. In this case, however, using the exact likelihood typically
requires numerical methods to evaluate integrals that can be evaluated
analytically under normality.

\begin{sloppypar}
We primarily focused on the evaluation of (objective) Bayesian properties of the
proposed methods. Further work is needed to evaluate their frequentist
properties, for example, with a carefully planned simulation study
\citep{Morris2019}. As in other recent studies \citep{Muradchanian2021,
  Freuli2022}, it would be interesting to simulate the realistic scenario of
questionable research practices and publication bias affecting the original
study to see how the adaptive downweighting of power priors can account for the
inflated original results.
\end{sloppypar}

\begin{appendices}
\section{Posterior distribution under the hierarchical model}
\label{app:postHierarch}

Under the hierarchical model from~\eqref{eq:hierarch-model}, the joint posterior
conditional on a heterogeneity $\tau^2$ is given by
\begin{align}
  \label{eq:jointpost2}
  f(\theta_{r}, \theta_{o}, \theta_{*} \given \that_{o}, \that_{r}, \tau^{2})
  = \frac{\prod_{i \in \{o, r\}} \Nor(\that_{i} \given \theta_{i}, \sigma^{2}_{i})
  \, \Nor(\theta_{i} \given \theta_{*}, \tau^{2}) \, k}{f(\that_{o}, \that_{r} \given \tau^2)}
\end{align}
with normalizing constant
\begin{align}
  f(\that_{o}, \that_{r} \given \tau^2)
  &= \int \prod_{i \in \{o, r\}} \Nor(\that_{i} \given \theta_i, \sigma^{2}_{i})
  \, \Nor(\theta_{i} \given \theta_{*}, \tau^{2}) \, k \,
  \text{d}\theta_o \,\text{d}\theta_r \,\text{d}\theta_* \nonumber \\
  &= \int \prod_{i \in \{o, r\}} \Nor(\that_{i} \given \theta_{*}, \sigma^{2}_{i} + \tau^2)
  k \, \text{d}\theta_* \nonumber \\
  &= k \, \Nor(\that_r \given \that_o, \sigma^{2}_{o} + \sigma^2_r + 2\tau^2).
  \label{eq:normConstHierarch}
\end{align}
To obtain the marginal posterior distribution of the replication effect size
$\theta_{r}$ we need to integrate out $\theta_{o}$ and $\theta_{*}$
from~\eqref{eq:jointpost2}. This leads to
\begin{align*}
  f(\theta_{r} \given \that_{o}, \that_{r}, \tau^{2})
  &= \frac{\int \prod_{i \in \{o, r\}} \Nor(\that_{i} \given \theta_{i}, \sigma^{2}_{i})
  \, \Nor(\theta_{i} \given \theta_{*}, \tau^{2}) \, k \,
  \text{d}\theta_o \,\text{d}\theta_*}{f(\that_{o}, \that_{r} \given \tau^2)} \\
  &= \frac{\Nor(\that_{r} \given \theta_{r}, \sigma^{2}_{r})
  \int \Nor(\theta_r \given \theta_{*}, \tau^2) \,
  \Nor(\that_{o} \given \theta_{*}, \sigma^{2}_{o} + \tau^2) \,
  \text{d}\theta_*}{\Nor(\that_r \given \that_o, \sigma^{2}_{o} + \sigma^2_r + 2\tau^2)} \\
  &= \frac{\Nor(\that_{r} \given \theta_{r}, \sigma^{2}_{r}) \,
  \Nor(\theta_r \given \that_o, \sigma^2_o + 2\tau^2)}{\Nor(\that_r \given \that_o, \sigma^{2}_{o} + \sigma^2_r + 2\tau^2)}
\end{align*}
which can be further simplified to identify the posterior given
in~\eqref{eq:posthierarch}.

When the heterogeneity $\tau^2$ is also assigned a prior distribution, the
posterior distribution can be factorized in the posterior conditional on
$\tau^2$ from~\eqref{eq:jointpost2} and the marginal posterior of $\tau^2$
\begin{align*}
  f(\tau^2, \theta_{r}, \theta_{o}, \theta_{*} \given \that_{o}, \that_{r})
  = f(\theta_{r}, \theta_{o}, \theta_{*} \given \that_{o}, \that_{r}, \tau^{2}) \,
  f(\tau^2 \given \that_{o}, \that_{r}).
\end{align*}
Integrating out $\theta_{r}, \theta_{o}$, and $\theta_{*}$ from the joint
posterior and using the previous results~\eqref{eq:normConstHierarch}, the
marginal posterior of $\tau^2$ can be derived to be
\begin{align*}
  f(\tau^2 \given \that_{o}, \that_{r})
  &= \frac{\int \prod_{i \in \{o, r\}} \Nor(\that_{i} \given \theta_{i}, \sigma^{2}_{i})
  \, \Nor(\theta_{i} \given \theta_{*}, \tau^{2}) \, k \, f(\tau^2) \,
  \text{d}\theta_o \,\text{d}\theta_r \,\text{d}\theta_*}{f(\that_o, \that_r)} \\
  &= \frac{f(\that_r, \that_o \given \tau^2)
  \, f(\tau^2)}{\int f(\that_r, \that_o \given \tau^2) \, f(\tau^2) \, \text{d}\tau^2} \\
  &= \frac{\Nor(\that_r \given \that_o, \sigma^{2}_{o} + \sigma^2_r + 2\tau^2)
  \, f(\tau^2)}{\int \Nor(\that_r \given \that_o, \sigma^{2}_{o} + \sigma^2_r + 2\tau^2) \, f(\tau^2) \, \text{d}\tau^2}.
\end{align*}

\section{Conditions for matching posteriors}
\label{app:mapping}

For the marginal posteriors of $\theta_r$ and $\theta$ to match it must hold for
every $\theta$ = $\theta_{r}$ that
\begin{align}
  f(\theta_r \given \that_{o}, \that_{r})
  &= f(\theta \given \that_{o}, \that_{r}) \nonumber \\
  \label{eq:margequal}
  \int_{0}^{\infty} f(\theta_{r} \given \that_{o}, \that_{r}, \tau^{2})\,
  f(\tau^{2} \given \that_{o}, \that_{r})\, \text{d} \tau^{2}
  &= \int_{0}^{1} f(\theta \given \that_{o}, \that_{r}, \alpha)\,
    f(\alpha \given \that_{o}, \that_{r}) \, \text{d} \alpha.
\end{align}
By applying a change of variables~\eqref{eq:tau2alpha} or~\eqref{eq:alpha2tau}
to the left or right hand side of \eqref{eq:margequal}, the marginal posteriors
conditional on $\tau^{2}$ and $\alpha$ match. It is now left to investigate
whether there are priors for $\tau^2$ and $\alpha$ so that also the marginal
posteriors of $\tau^{2}$ and $\alpha$ match. The marginal posterior distribution
of $\alpha$ is proportional to
\begin{align*}
  f(\alpha \given \that_{o}, \that_{r})
  \propto f_\alpha(\alpha) \,
  \Nor(\that_{r}\given \that_{o}, \sigma^{2}_{r} + \sigma^{2}_{o}/\alpha).
\end{align*}
After a change of variables $\tau^{2} = (1/\alpha - 1)\,(\sigma^{2}_{o}/2)$ the
marginal posterior becomes
\begin{align*}
  f(\tau^{2} \given \that_{o}, \that_{r})
  \propto f_\alpha\left(\frac{\sigma^{2}_{o}}{2 \tau^{2} + \sigma^{2}_{o}}\right) \,
  \frac{2\sigma^{2}_{o}}{(2 \tau^{2} + \sigma^{2}_{o})^{2}} \,
  \Nor(\that_{r}\given \that_{o}, \sigma^{2}_{r} + \sigma^{2}_{o} + 2 \tau^{2}),
\end{align*}
Since, as shown in Appendix~\ref{app:postHierarch}, the marginal posterior of
$\tau^2$ under the hierarchical model is proportional to
\begin{align*}
  \label{eq:margposttau2}
  f(\tau^{2} \given \that_{o}, \that_{r})
  \propto
  f_{\tau^2}(\tau^{2}) \,
\Nor(\that_{r}\given \that_{o}, \sigma^{2}_{r} + \sigma^{2}_{o} + 2 \tau^{2}),
\end{align*}
the marginal posteriors of the effect sizes $\theta$ and $\theta_{r}$ match if
\begin{align*}
    f_{\tau^2}(\tau^{2}) = f_\alpha\left(\frac{\sigma^{2}_{o}}{2 \tau^{2} + \sigma^{2}_{o}}\right) \,
  \frac{2\sigma^{2}_{o}}{(2 \tau^{2} + \sigma^{2}_{o})^{2}}
\end{align*}
holds for every $\tau^{2}\geq 0$.

\section{The generalized beta and F distributions}
\label{app:distribution}

A random variable $X \sim \mbox{GBe}(a, b, \lambda)$ with density function
\begin{align}
    f(x\given a, b, \lambda)
    = \frac{\lambda^a \, x^{a - 1} \, (1 - x)^{b - 1}}{\mbox{B}(a, b) \,
    \{1 - (1 - \lambda)x\}^{a + b}} \, \mathbf{1}_{[0, 1]}(x)
\end{align}
follows a generalized beta distribution \citep[in the parametrization
of][]{Libby1982} with $\mathbf{1}_{S}(x)$ denoting the indicator function that
$x$ is in the set $S$. A random variable $X \sim \mbox{GF}(a, b, \lambda)$ with
density function
\begin{align}
    f(x\given a, b, \lambda)
    = \frac{\lambda^a \, x^{a - 1}}{\mbox{B}(a, b) \,
    (1 + \lambda x)^{a + b}} \, \mathbf{1}_{[0, \infty)}(x)
\end{align}
follows a generalized F distribution \citep[in the parametrization
of][]{PhamGia1989}.

\end{appendices}

\section*{Software and data}
The CC-By Attribution 4.0 International licensed data were downloaded
from \url{https://osf.io/42ef9/}. All analyses were conducted in the R
programming language version 4.3.1 \citep{R}. The code and data to reproduce this manuscript is available
at \url{https://github.com/SamCH93/ppReplication}. A snapshot of the GitHub
repository at the time of writing this article is archived
at \url{https://doi.org/10.5281/zenodo.6940237}. We also provide an R package
for estimation and testing under the power prior framework
(\url{https://CRAN.R-project.org/package=ppRep}). The package can be installed
by running \texttt{install.packages("ppRep")} from an R console.

\section*{Acknowledgments}
We thank \citet{Protzko2020} for publicly sharing their data. We thank
Małgorzata Roos for helpful comments on a draft of the manuscript. We thank the
associate editor and the two anonymous reviewers for many excellent comments and
suggestions. This work was supported in part by an NWO Vici grant
(016.Vici.170.083) to EJW, an Advanced ERC grant (743086 UNIFY) to EJW, and a
Swiss National Science Foundation mobility grant (189295) to LH and SP.

\section*{Conflict of interest}
The authors have no conflicts of interest to declare.

\bibliographystyle{apalikedoiurl}
\bibliography{bibliography}

\end{document}